# HIGH FIELD PERFORMANCE OF NANO-DIAMOND DOPED $MgB_2$ SUPERCONDUCTOR

Arpita Vajpayee, V.P.S. Awana, and H. Kishan
National Physical Laboratory, Dr. K.S. Krishnan Marg, New Delhi-110012, India

A.V. Narlikar
UGC-DAE Consortium for Scientific Research, University Campus, Khandwa Road, Indore-452017, India

G.L. Bhalla
Department of Physics and Astrophysics, Delhi University, New Delhi, India

X.L. Wang
ISEM, University of Wollongong, NSW 2522 Australia

Polycrystalline $MgB_2$-$n$$D_x$ (x= 0 to 0.1) samples are synthesized by solid-state route with ingredients of Mg, B and *n*-Diamond. The results from magneto-transport and magnetization of *nano*-diamond doped $MgB_2$-$n$$D_x$ are reported. Superconducting transition temperature ($T_c$) is not affected significantly by x up to x = 0.05 and latter decreases slightly for higher x > 0.05. $R(T)$ vs $H$ measurements show higher $T_c$ values under same applied magnetic fields for the *nano*-diamond added samples, resulting in higher estimated $H_{c2}$ values. From the magnetization measurements it was found that irreversibility field value $H_{irr}$ for the pristine sample is 7.5 Tesla at 4 K and the same is increased to 13.5 Tesla for 3-wt% *n*D added sample at the same temperature. The $J_c(H)$ plots at all temperatures show that $J_c$ value is lowest at all applied fields for pristine $MgB_2$ and the sample doped with 3-wt% *n*D gives the best $J_c$ values at all fields. For the pure sample the value of $J_c$ is of the order of $10^5$ A/cm$^2$ at lower fields but it decreases very fast as the magnetic field is applied and becomes negligible above 7 Tesla. The $J_c$ is 40 times higher than pure $MgB_2$ at 10 K at 6 Tesla field in case of 3%*n*D doped sample and its value is still of the order of $10^3$ A/cm$^2$ at 10 Tesla for the same sample. On the other hand at 20K the 5%$n$D sample shows the best performance at higher fields. These results are discussed in terms of extrinsic pinning due to dispersed $n$-Diamond in the host $MgB_2$ matrix along with the intrinsic pinning due to possible substitution of C at Boron site and increased inter-band scattering for highly doped samples resulting in extraordinary performance of the doped system.

## I. INTRODUCTION

The improvement of the pinning behavior in $MgB_2$ is being considered as a key issue in the superconductivity field; many attempts have been made in this direction for enhancing the pinning strength of $MgB_2$. In the last five years, various carbon sources [1-3], silicates [4], single elements [5,6], carbohydrates [7], organic acids [8] and other compounds [9,10] have been doped in $MgB_2$ and reported. It has become more clear that structural defects with dimensions at nano scale play an important role for flux pinning i.e. nanoparticle doping is the most effective approach to improve the superconducting performance of $MgB_2$ and making $MgB_2$ acceptable for practical applications [11]. Nano particle doping can modify structure and electronic properties such as $T_c$, upper critical field and its anisotropy, gaps width, inter and intra-band scattering, defect structure etc. [12-14]. In present paper, we are discussing the effect of *nano* diamond addition on polycrystalline $MgB_2$ i.e. how the superconducting performance of pristine sample is modified by addition of $n$-Diamond.

## II. EXPERIMENTAL

The $MgB_2$-*nano* Diamond composites with composition $MgB_2$ - $nD_x$ (x= 0%, 1%, 3%, 5%, 7% & 10%) are synthesized by solid-state reaction route with ingredients of Mg, B and $n$-Diamond. The Mg powder used is from *Reidel-de-Haen* of assay 99%, B powder is amorphous and *Fluka* make of assay 95-97%. The $n$-Diamond powder is from Aldrich with average particle size (*APS*) of 7-10 nm. For synthesis of $MgB_2$-$nD_x$ samples, the nominal weighed samples are ground thoroughly, palletized, encapsulated in soft iron tube and put in a quartz tube inserted in a programmable furnace under flow of argon at ambient pressure. The temperature of furnace is programmed to reach 850 $^0$C over 2 hours, hold at same temperature for two and half hours, and subsequently cooled to room temperature over a span of 6 hours in the same Argon atmosphere. The x-ray diffraction pattern of the compound was recorded with a diffractometer using $CuK_\alpha$ radiation. SEM studies were carried out on these samples using a Leo 440 (Oxford Microscopy: UK) instrument. The magnetoresistivity, *ρ(T)H*, was measured with *H* applied perpendicular to current direction, using four-probe technique on *Quantum Design PPMS*. Magnetization measurements are carried out with a Quantum-Design *SQUID* magnetometer *MPMS*-7.

## III. RESULTS AND DISCUSSIONS

X-ray diffractions patterns show that all the samples of $MgB_2$-*n*$D_x$ series are of nearly single phase with only a small quantity of un-reacted MgO in them. For highest *n*$D_x$ sample, i.e., x = 10wt%, the main intensity peak is broadened slightly (patterns not shown). The lattice parameters remain nearly invariant for all the samples until x = 5wt% for $MgB_2$-*n*$D_x$. For 10 wt% sample, slight decrease in '*a*' is seen. It means for this higher concentration some diamond particles has got broken into carbon atoms, which substituted at Boron site resulting in slight decrease in 'a' but at low concentration there is not such indication of breakage of *n*-Diamond.

Variation of resistance with temperature under applied magnetic fields up to 8 Tesla *R(T)H* is shown in fig.1 (only transition zone). As we can see from this figure with increasing *n*-Diamond concentration, the superconducting onset temperature remains unchanged and the transition width changes slightly. This indicates that no substitution has taken place in this system. It is noted that *R(T)H* curves for the doped samples shifted with increasing field much more slowly than the undoped one. For pure $MgB_2$ the $T_c(R = 0)$ value is observed at 38 and 18 K in zero and 8 Tesla applied fields respectively. Interestingly, the $T_c(R = 0)$ for $MgB_2$-5wt% *n*D sample in comparison to pure $MgB_2$ is lowered to 36 K in zero applied field, the same is increased to 22 K under 8 Tesla field.

Now, to see the effect of *n*-Diamond addition on the upper critical field $H_{c2}$, using resistivity variation under magnetic field plots we estimated the $H_{c2}$ values i.e. $H_{c2}$ obtained from the 90% values of the resistive transitions. The variation of $H_{c2}$ with respect to temperature is not shown here (see Ref. 15), which demonstrates that on increasing the concentration of *n*-Diamond a clear shift in upper direction can be achieved. The behavior of 3-wt% & 5-wt% *n*-Diamond added samples is quite competitive; but since the $T_c$ of 5%*n*D is lower than 3%*n*D so we can say that 5%*n*D sample shows better improvement in $H_{c2}$ value. The formation of nanodomains structures is due to the presence of dispersed diamond in parent $MgB_2$ grains (we have reported the HRTEM and ED pattern in another paper [15]) causing the disorder in lattice. These nanodomains, which are comparable in size (~ 8 to 10 nm) with the coherence length of $MgB_2$ could result in strong in-plane and out-of-plane scattering and contribute to the increase of $H_{c2}$ value in a wide temperature regime.

In fig.2 the magnetic hysteresis loop for all the doped samples (x = 0%, 1%, 3%, 5%, 7% & 10%) at 10K is shown in applied field up to 13 Tesla. This tells that the closing of hysteresis loop for pure $MgB_2$ is at ~ 6.5 Tesla whereas the same is opened up to ~11 Tesla for 3-wt% doped sample. This indicates that there will be significant improvement in irreversibility field values for the doped samples. The irreversibility fields ($H_{irr}$) is the one at which the magnetic hysteresis loop is nearly closed with $M \approx 0$. The $H_{irr}$ versus $x$ (concentration on *nano* diamond) plot is shown in inset of fig.2 at 4k, 10K & 20K. The best values of $H_{irr}$ among these diamond added samples ($MgB_2$-$n$$D_x$) is found for the sample with x=3%. The $H_{irr}$ values for the undoped samples are 7.5, 6.5 & 4.5 Tesla at 4K, 10K & 20K respectively; the same is increased up to 13.5, 11.3 & 5.5 Tesla for the 3-wt% *n*-Diamond added sample, reveals that nano diamond doping significantly improved the $H_{irr}$ values. These results show that diamond doping enhances the flux pinning strength in parent $MgB_2$ significantly. These results are in well agreement with previously reported data [16].

Using *M(H)* loop we estimated the critical current density by invoking Bean's critical state formula. Fig. 3 shows the magnetic critical current density ($J_c$) vs applied field (*H*) at 20K & 10K. For all the samples (doped and undoped) the $J_c$ attains the value greater than $10^5$ A/cm$^2$ at low fields at both the temperatures. At 20K $J_c$ falls to 100 A/cm$^2$ at 4.5 Tesla for pure $MgB_2$ whereas for rest of the samples this $J_c$ value is reached only at nearly 6 Tesla. At 20 K the $J_c$ is improved by an order of magnitude for doped samples as compared to the pure $MgB_2$ at high fields but for doped samples all the plots are merging nearly at the same field (6 Tesla) i.e. not much difference in $J_c(H)$ behavior is seen for doped samples at higher fields. For 10K the situation is quite different at higher fields i.e. the rate of $J_c$ drop is smaller than for all samples compared to undoped $MgB_2$. The $J_c$ is 40 times higher than pure $MgB_2$ at 10K at 6 Tesla applied field in case of 3-wt% *n*-Diamond added sample and $J_c$ value is still of the order of $10^3$ A/cm$^2$ at 10 Tesla for the same sample. These results are quite comparable as reported earlier [17]. The

better $J_c(H)$ performance of *n*-Diamond added samples indicates towards that the additional impurities at nanoscale introduced by *n*-Diamond serve as strong pinning centers to improve flux pinning in higher applied fields.

To confirm the improved flux pinning behavior through diamond doping, the field dependence of normalized flux pinning force ($F_p$ / $F_{p,\,max}$) is shown in fig. 4(a) & (b) at 20K & 10K. The relationship between flux pinning force and critical current density could be described by [18]

$$F_p = \mu_0 \, J_c(H) \, H \tag{1}$$

Where $\mu_0$ is the magnetic permeability in vacuum. These figures depict that the pinning forces are significantly improved than the pure $MgB_2$ above 2 Tesla for *n*-Diamond doped samples at both the temperatures because curves for the doped samples are shifted towards the higher fields, indicating enhanced flux pinning force in high fields. As revealed by the microstructure analysis [15] *nano* diamond particles of size around ~ 10 nm are dispersed in $MgB_2$ matrix. Because these particles have the size comparable to the coherence length of $MgB_2$, It is highly possible that they may work as point pinning centers, causing a shift of the curve in $F_p$ / $F_{p,max}$ vs $H$ curve towards the higher field. It can be seen from fig. 4(a) & (b) that the peak corresponding to 3-wt% *n*-Diamond doped sample is much broader than those of other samples, indicating highest pinning strength in this sample at higher fields; which is consistent with the results of $J_c$ (fig. 3).

## IV. CONCLUSIONS

In summary, the effect of *nano* diamond doping on critical temperature ($T_c$), critical current density ($J_c$) and flux pinning was investigated under a wide range of magnetic field. It was found that the *n*-Diamond addition enhances the flux pinning with modest reduction of the $T_c$. $J_c$ was enhanced by a factor of 40 in comparison to undoped sample for 3-wt% *n*-Diamond added sample at 10K at 6 Tesla. The $H_{irr}$ values for the undoped samples are 7.5, 6.5 & 4.5 Tesla at 4K, 10K & 20K respectively; the same is increased up to 13.5, 11.3 & 5.5 Tesla for the 3-wt% *n*-Diamond added sample. The highly dispersed nano diamond particles are acting as strong pinning centers and are responsible for better performance of doped samples. The present work shows that *nano* diamond doping is a promising way to fabricate high performance $MgB_2$ bulk material with excellent values of $J_c(H)$, $H_{c2}$ & $H_{irr}(T)$.

**ACKNOWLEDGEMENT**

Dr. Rajeev Rawat from *CSR*-Indore is acknowledged for the resistivity under magnetic field measurements. Mr. Kranti Kumar and Dr. A. Banerjee are acknowledged for the high field magnetization measurements. Further *DST*, Government of India is acknowledged for funding the 14 Tesla-PPMS-VSM at CSR, Indore. The authors from NPL would like to thank Dr. Vikram Kumar (DNPL) for his great interest in present.

**FIGURE CAPTIONS**

Figure1. Superconducting transition zone of Resistance vs Temperature plot under applied magnetic field *R(T)H* up to 8 Tesla for pure, 5-wt% & 10-wt% *nano* diamond added samples

Figure 2. Magnetization loop *M(H)* at 10K for $MgB_2$-*n*$D_x$ (x=0%, 1%, 3%, 5%, 7% & 10%) up to 12 Tesla field and inset shows the behavior of irreversibility field $H_{irr}$ with respect to *n*-Diamond concentration at 4K, 10K & 20K

Figure 3. Critical current density ($J_c$) variation with respect to applied magnetic field (*H*) (a) at 20K & (b) at 10K

Figure 4. Variation of reduced flux pinning force ($F_p/F_{p,max}$) with magnetic field for $MgB_2$-*n*$D_x$ (x=0%, 1%, 3%, 5% & 10%) at 20K & 10K

**Fig.1**

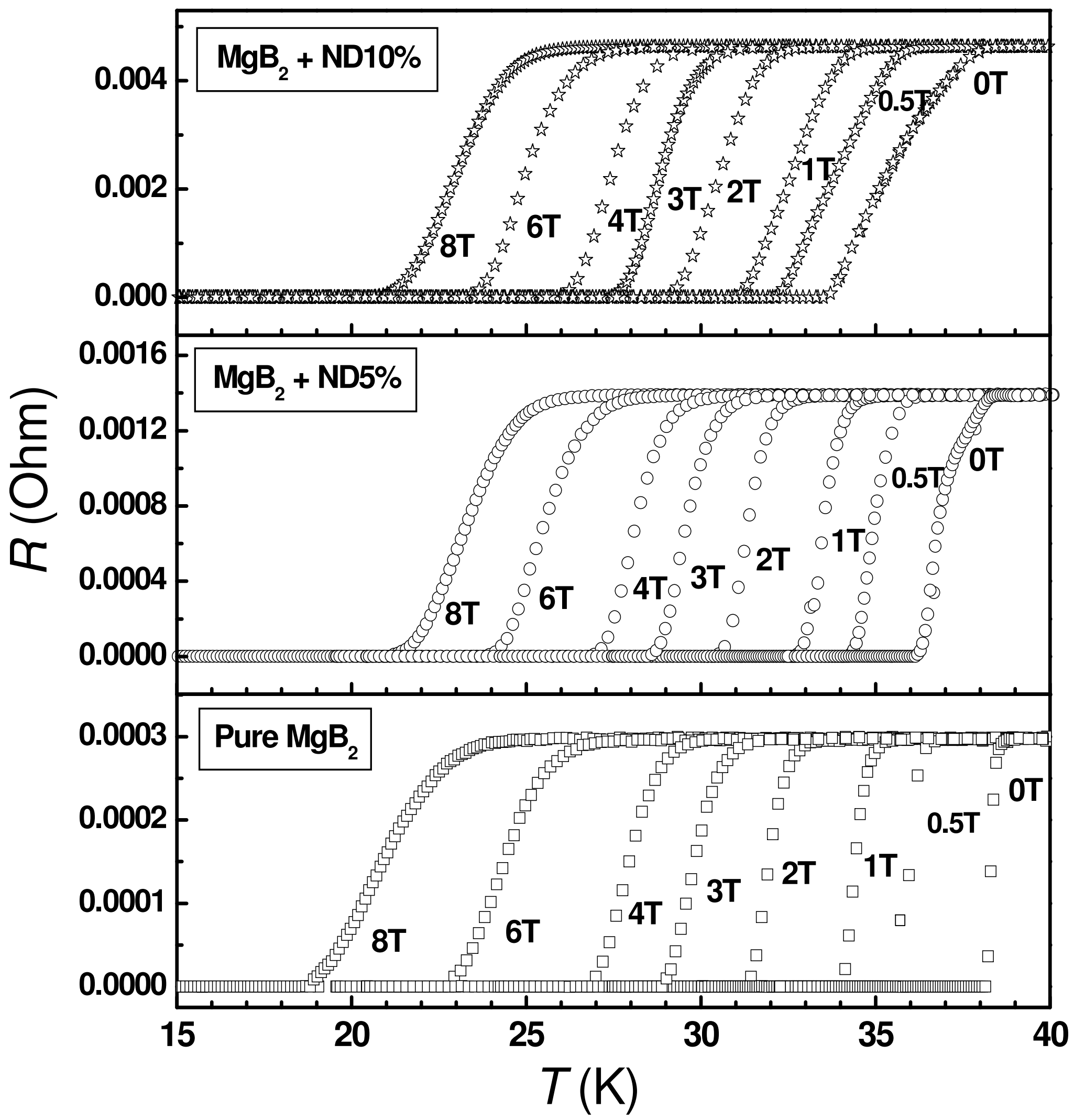

**Fig. 2**

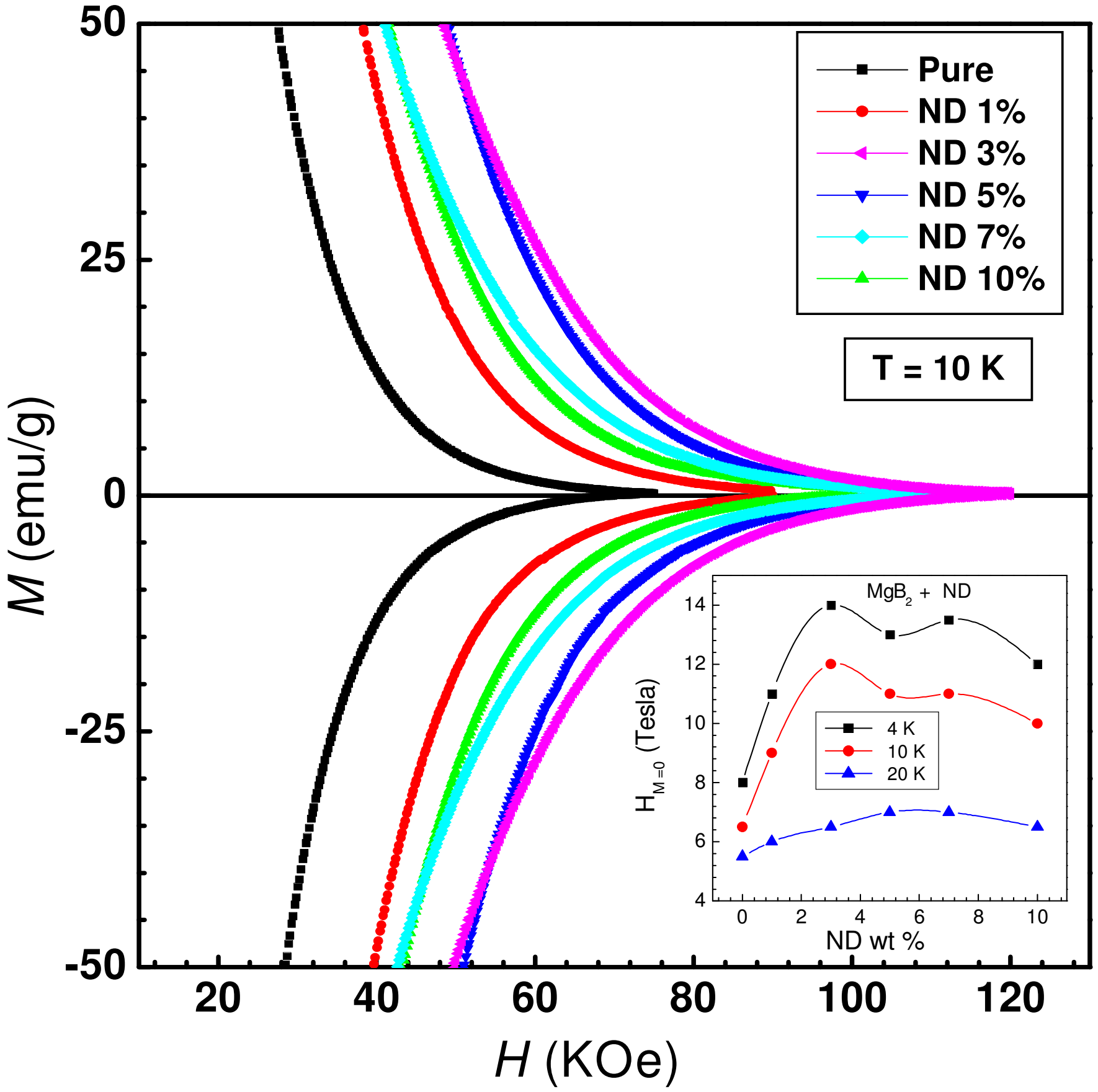

**Fig. 3(a) & 3(b)**

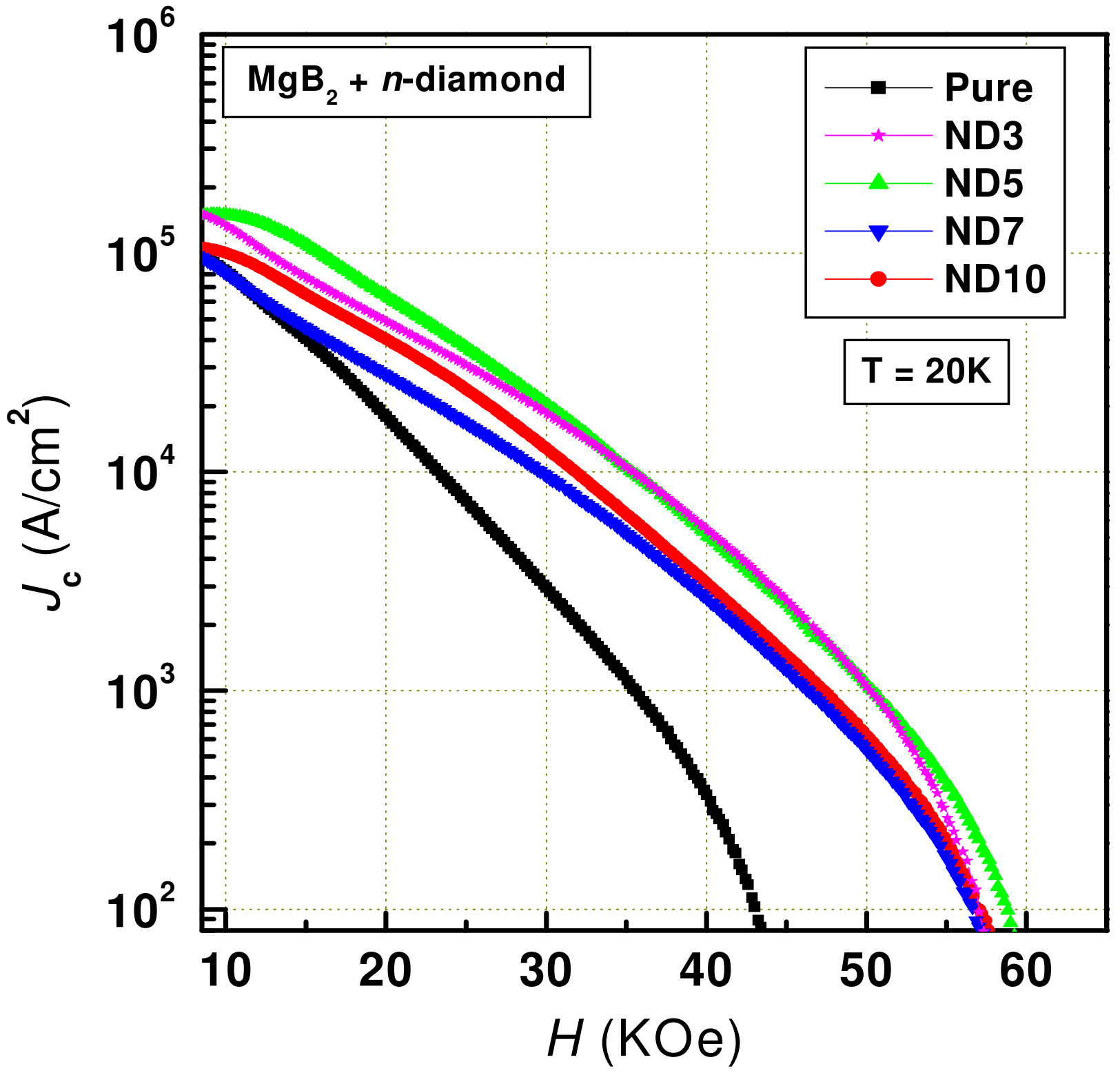

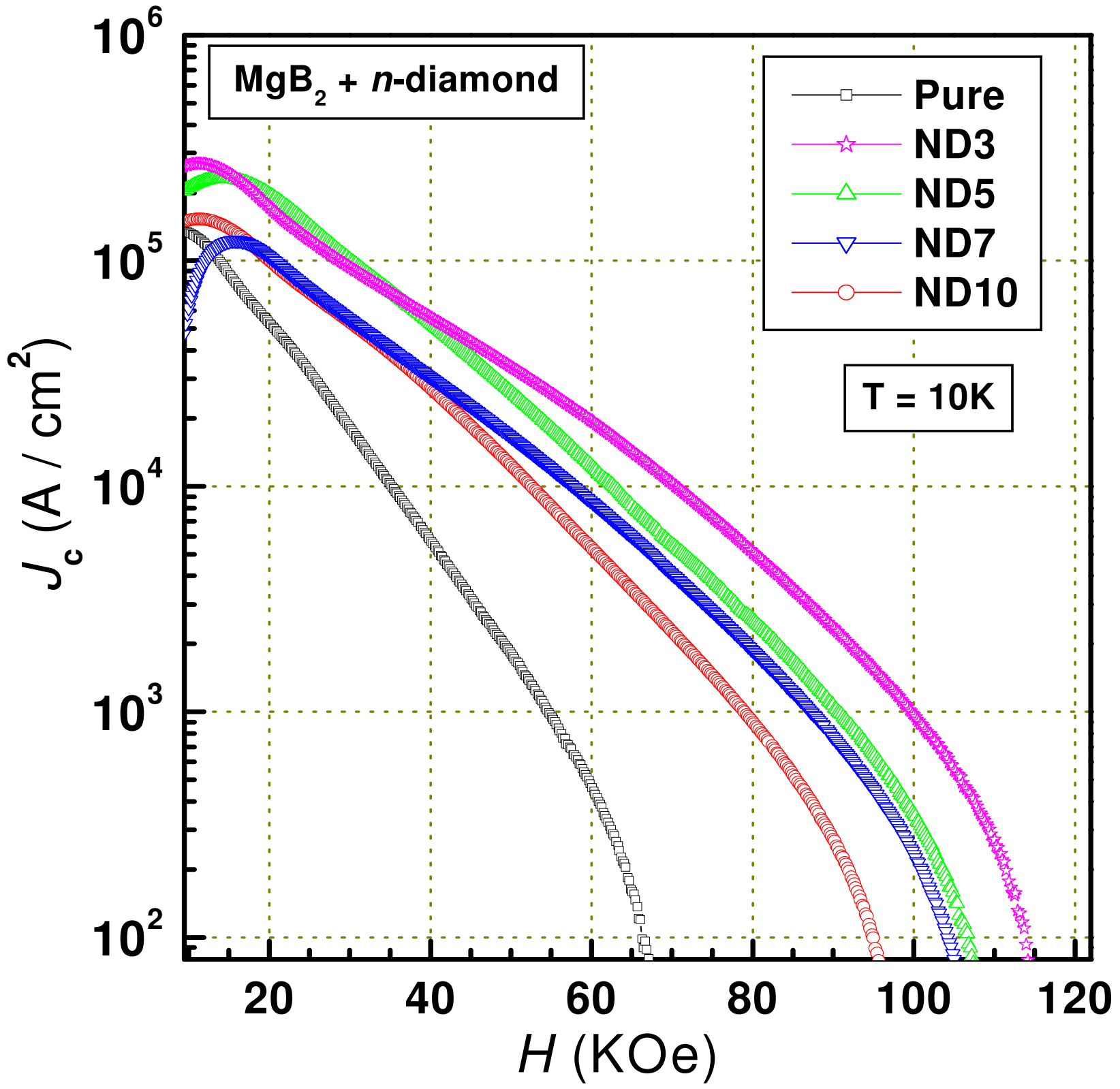

**Fig. 4(a) & 4(b)**

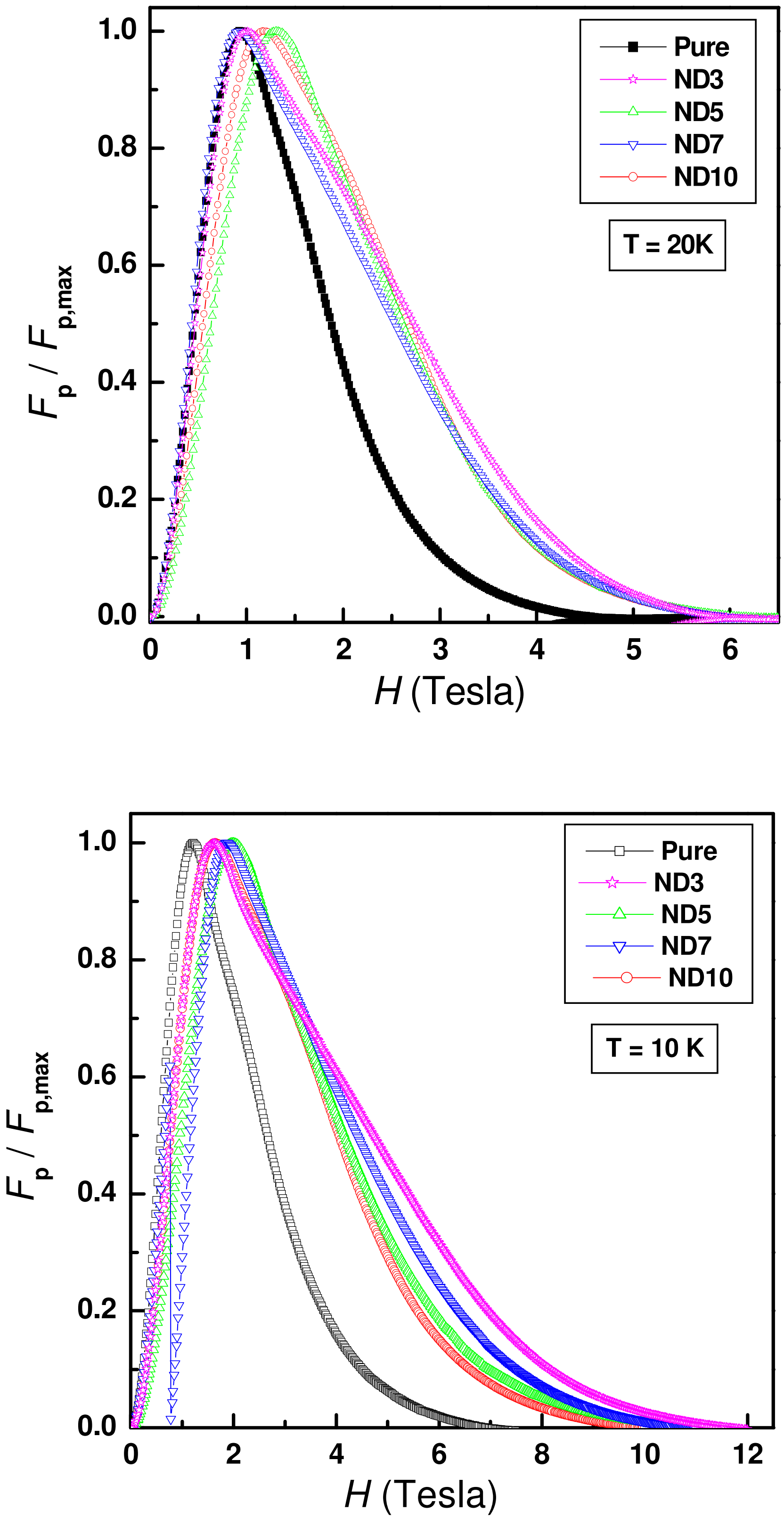